\documentclass[aps,cpl,reprint,superscriptaddress]{revtex4-1}
\usepackage{amsmath}
\usepackage{graphicx}
\usepackage{longtable}
\usepackage{hyperref}

\begin{document}

\title{Seeking for reliable double-hybrid density functionals without fitting parameters: The PBE0-2 functional}

\date{\today}

\author{Jeng-Da Chai}
\email[Author to whom correspondence should be addressed. Electronic mail: ]{jdchai@phys.ntu.edu.tw.} 
\affiliation{Department of Physics, National Taiwan University, Taipei 10617, Taiwan} 
\affiliation{Center for Theoretical Sciences and Center for Quantum Science and Engineering, National Taiwan University, Taipei 10617, Taiwan} 

\author{Shan-Ping Mao} 
\affiliation{Department of Physics, National Taiwan University, Taipei 10617, Taiwan}

\begin{abstract}

Without the use of any empirical fitting to experimental or high-level {\it ab initio} data, we present a double-hybrid density functional approximation for the exchange-correlation energy, 
combining the exact Hartree-Fock exchange and second-order M\o ller-Plesset (MP2) correlation with the Perdew-Burke-Ernzerhof (PBE) functional. This functional, denoted as PBE0-2, is shown to be accurate 
for a wide range of applications, when compared with other functionals and the {\it ab initio} MP2 method. The qualitative failures of conventional density functional approximations, such as self-interaction 
error and noncovalent interaction error, are significantly reduced by PBE0-2. 

\end{abstract}

\maketitle

\section{Introduction}

Over the past two decades, Kohn-Sham density functional theory (KS-DFT) \cite{KS1965} has become a very popular electronic structure method for molecular and solid-state systems, due to its 
computational efficiency and reasonable accuracy. Although the exact exchange-correlation (XC) density functional $E_{xc}[\rho]$ in KS-DFT remains unknown, accurate approximations to $E_{xc}[\rho]$ have been 
continuously developed to extend the applicability of KS-DFT to a wide range of systems \cite{DFTreview}. 

The hierarchy of approximations to $E_{xc}[\rho]$ has been formulated as a {\it Jacob's ladder} connecting the earth (Hartree theory) to the heaven (chemical accuracy) \cite{constraint}. 
The first rung of the ladder is the local density approximation (LDA) \cite{LDAX,LDAC}, representing the XC energy density by the local density. Going beyond the LDA, there has been an increasing 
interest in the ``parameter-free" functionals developed by Perdew and co-workers, such as PBE \cite{PB1996} and TPSS \cite{TP2003}, demonstrating the usefulness of these functionals. 
As LDA, PBE, and TPSS belong to the first, second, and third rungs of the ladder, respectively, they systematically improve upon the description of short-range XC effects by climbing up the ladder 
(with slightly increasing computational costs). However, due to the lack of accurate treatment of nonlocal XC effects, these semilocal functionals (first three rungs) can lead to significant errors, characterized by 
self-interaction error (SIE), noncovalent interaction error (NCIE), and static correlation error (SCE), in situations where these failures occur \cite{DFTreview,Chai2012}. 

The SIEs of density functional approximations (DFAs) can be reduced by hybrid DFT methods (fourth rung) \cite{B1993_2}, combining a fraction of the exact Hartree-Fock (HF) exchange with a semilocal 
functional. As the fraction of HF exchange included in a hybrid functional can be rationalized by perturbation theory arguments \cite{DFA0}, the PBE-based hybrid functional PBE0 \cite{PBE0}, 
has gradually gained its popularity. In view of the successive evolution of the PBE-based functionals: LDA (first rung) $\rightarrow$ PBE (second rung) $\rightarrow$ TPSS (third rung) 
$\rightarrow$ PBE0 (fourth rung) $\rightarrow$ $?$ (fifth rung), naturally, the question is, what could be a functional on the fifth rung (highest rung) of the ladder? In this work, we propose a PBE-based 
double-hybrid functional (fifth rung), and demonstrate its superiority to the functionals on the first four rungs of the ladder, for a wide range of applications.

\section{The PBE0-2 Functional}

The SIEs and NCIEs associated with DFAs can be simultaneously reduced by double-hybrid methods (fifth rung) \cite{DH1,DH2,G2006,XYG3,CH2009,B2PLYPD3}, combining a fraction of HF exchange 
and a fraction of second-order M\o ller-Plesset (MP2) correlation \cite{MP1934} with a semilocal functional. Accordingly, a PBE-based double-hybrid functional is given by 
\begin{eqnarray}
	E_{xc} &=& a_x E_{x}^{\text{HF}} + (1 - a_x) E_{x}^{\text{PBE}} + (1 - a_c) E_{c}^{\text{PBE}} \nonumber \\
	             &+& a_c E_{c}^{\text{MP2}},                                
 \label{DH}
\end{eqnarray}
where $E_{x}^{\text{HF}}$ is the HF exchange, $E_{x}^{\text{PBE}}$ is the PBE exchange, $E_{c}^{\text{PBE}}$ is the PBE correlation, and $E_{c}^{\text{MP2}}$ is the MP2 correlation (a perturbative term 
evaluated with the orbitals obtained using the first three terms of the energy). In this work, the two mixing parameters, $a_x$ and $a_c$, are determined by physical arguments (i.e., without the use of any 
empirical fitting to experimental or high-level {\it ab initio} data). 

Based on the double-hybrid approximations proposed by Sharkas, Toulouse, and Savin \cite{ST2011}, and the closely related work of Br\' emond and Adamo \cite{BA2011}, 
Toulouse \textit{et al.} \cite{TS2011} have proposed the linearly scaled one-parameter double-hybrid (LS1DH) approximation (see Eq.\ (10) of Ref.\ \cite{TS2011}), showing the relationship between the mixing 
parameters, $a_{x}$ and $a_{c}$, in a double-hybrid functional, 
\begin{equation}
a_{c} = (a_{x})^{3}.
\label{LS1DH}
\end{equation}
Applying Eq.\ (\ref{LS1DH}) to Eq.\ (\ref{DH}), only one of the mixing parameters needs to be determined. 

Functionals on the first four rungs of the ladder fail to describe long-range van der Waals (vdW) interactions in regions where there is no overlap of the subsystems, due to the lack of long-range correlation 
effects \cite{DFTreview,Dobson2}. This problem can be greatly reduced by the DFT-D (KS-DFT with empirical dispersion corrections) schemes \cite{DFT-D1,DFT-D2,wB97X-D,B2PLYPD3} or by a fully nonlocal 
density functional for vdW interactions (vdW-DF) \cite{vdW}, showing generally acceptable performance on several noncovalent systems \cite{Sherrill}. 

By contrast, the MP2 correlation in a double-hybrid functional is itself fully nonlocal, and thus responsible for describing long-range vdW interactions. 
To determine the fraction $a_{c}$ of MP2 correlation in Eq.\ (\ref{DH}), consider the two limiting cases where $a_{c} = (a_{x})^3 = 0$ and $a_{c} = (a_{x})^3 = 1$, which reduce to PBE and MP2, respectively. 
In view of the underbinding tendency of PBE and overbinding tendency of MP2 in describing the interaction energies of many vdW complexes, a mixing of 50\% of PBE correlation with 50\% of MP2 correlation 
seems physically justified (as vdW interactions are inherently {\it correlation} effects). Employing this half-and-half mixing of the PBE and MP2 correlation functionals with the LS1DH approximation in Eq.\ (\ref{LS1DH}), 
we obtain 
\begin{equation}
a_{c} = 1/2, \ \  a_{x} = (1/2)^{1/3} \approx 0.793701. 
\label{PBE0-2}
\end{equation}
Our resulting PBE-based double-hybrid functional, defined in Eqs.\ (\ref{DH}) and (\ref{PBE0-2}), is denoted as PBE0-2, where the ``0" refers to no fitting parameters, and 
the ``-2" refers to the post-KS treatment for the MP2 correlation. 

By contrast, the mixing parameters $a_{x} = 1/2$ and $a_{c} = (1/2)^{3} = 1/8$, are adopted in PBE0-DH \cite{BA2011}, another PBE-based double-hybrid functional (also based on 
Eqs.\ (\ref{DH}) and (\ref{LS1DH})), wherein a half-and-half mixing of the PBE and HF exchange functionals is {\it effectively} employed. Therefore, a comparison between the performance of PBE0-2 and PBE0-DH 
is particularly important, as it shows the significance of the physical arguments used in determining the $a_{x}$ and $a_{c}$.

\section{Results}

For a comprehensive comparison of different functionals, we examine the performance of PBE0-2, other PBE-based functionals (LDA \cite{LDAX,LDAC}, PBE \cite{PB1996}, TPSS \cite{TP2003}, 
PBE0 \cite{PBE0}, and PBE0-DH \cite{BA2011}), three popular functionals (B3LYP (a hybrid functional) \cite{B1993_2,B3LYP}, B2PLYP (a double-hybrid functional) \cite{G2006}, and 
B2PLYP-D3 (a double-hybrid functional with empirical dispersion corrections) \cite{B2PLYPD3}), and the {\it ab initio} MP2 method \cite{MP1934}, 
on various test sets involving the 223 atomization energies (AEs) of the G3/99 set \cite{CR1997, CR1998, CR2000}, 
the 40 ionization potentials (IPs), 25 electron affinities (EAs), and 8 proton affinities (PAs) of the G2-1 set \cite{PH1989}, the 76 barrier heights of the NHTBH38/04 and HTBH38/04 sets \cite{ZL2004_2, ZG2005}, 
the 22 noncovalent interactions of the S22 set \cite{JS2006, TH2010}, the 66 noncovalent interactions of the S66 set \cite{RL2011}, two interaction energy curves of weakly bound complexes, 
isodesmic reaction energies, and two dissociation curves of symmetric radical cations, with a development version of \textsf{Q-Chem 3.2} \cite{SF2006}. Detailed information about some of the test sets can 
be found in Ref.\ \cite{wB97}. The error for each entry is defined as (error = theoretical value $-$ reference value). 
Spin-restricted theory is used for singlet states and spin-unrestricted theory for others. Results are computed using the 6-311++G(3df,3pd) basis set, unless noted otherwise. 
For efficiency, the resolution-of-identity (RI) approximation \cite{KF1997} is used for calculations with the MP2 correlation (using sufficiently large auxiliary basis sets). The interaction energies of 
the weakly bound systems are computed with the counterpoise correction \cite{BB1970} for the basis set superposition errors (BSSE). 

As shown in Table \ref{tab1}, for the AEs of the G3/99 set, B2PLYP and B2PLYP-D3 have the best performance (partly due to the fact that this test set overlaps with their training sets). 
PBE0-2 performs similarly to PBE0-DH, PBE0, TPSS, and B3LYP, and significantly outperforms LDA, PBE, and MP2. 
For the IPs, EAs, and PAs of the G2-1 set, LDA performs poorly, while all the other functionals perform similarly to MP2. 
For the barrier heights of the NHTBH38/04 and HTBH38/04 sets, PBE0-2, PBE0-DH, B2PLYP, and B2PLYP-D3 have comparable performance, and are the best functionals here \cite{supp}. 
For the noncovalent interactions of the S22 and S66 sets, B2PLYP-D3 performs the best (mainly due to the empirical dispersion corrections). PBE0-2 performs similarly to MP2, and significantly 
outperforms all the other functionals \cite{supp}. 

\begin{figure}
	\includegraphics[scale=0.4]{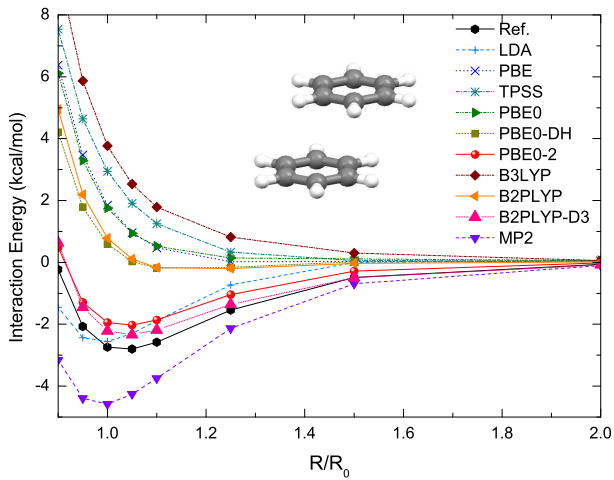}
	\caption{\label{benzene_pi} Interaction energy curve for the parallel-displaced configuration of the benzene dimer as a function of the intermonomer distance $R$ (defined in Ref.\ \cite{RL2011}), 
	where $R_0$ is the equilibrium distance.}
\end{figure}

\begin{figure}
	\includegraphics[scale=0.4]{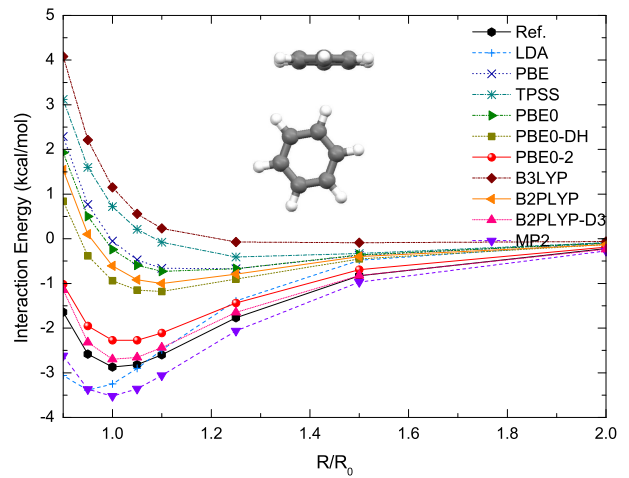}
	\caption{\label{benzene_t} Same as Fig.\ {\ref{benzene_pi}}, but for the T-shaped configuration.}
\end{figure}

Here, we calculate the interaction energy curves for the parallel-displaced (in Fig.\ \ref{benzene_pi}) and T-shaped (in Fig.\ \ref{benzene_t}) configurations of the benzene dimer as functions of the intermonomer 
distance $R$ (defined in Ref.\ \cite{RL2011}), where the optimized geometries and reference values are taken from the S66$\times8$ set \cite{RL2011}. 
Overall, the predicted interaction energy curves of B2PLYP-D3 and PBE0-2 are the closest ones to the reference curves (within an error of 1 kcal/mol). 
LDA predicts too short equilibrium distances, and all the other functionals predict severely underbinding or even repulsive curves. 
MP2 incorrectly predicts the parallel-displaced configuration to be more stable than the T-shaped one. 

\begin{figure}
	\includegraphics[scale=0.4]{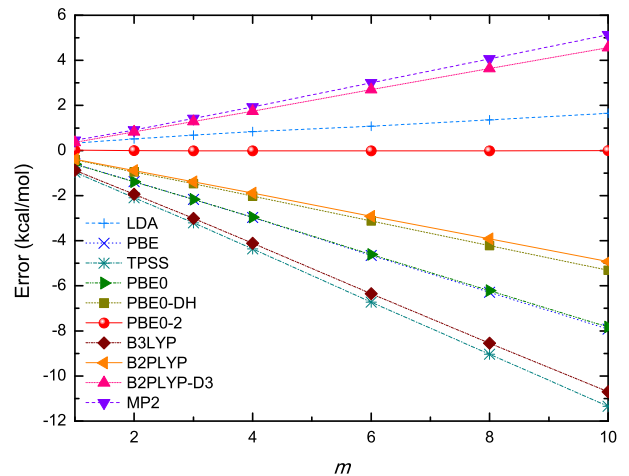}
	\caption{\label{Alkane} Errors for isodesmic reaction energies of \textit{n}-alkanes to ethane.}
\end{figure}

The isodesmic reaction energies of \textit{n}-alkanes to ethane have been known to give systematic errors in conventional DFT calculations \cite{RZ2000,SE1,ST2010,G2010,SE2}. 
The considered bond separation reaction, 
\begin{align}
	\text{CH}_{3}(\text{CH}_{2} )_{m}\text{CH}_{3}+m\text{CH}_{4} \rightarrow (m+1)\text{C}_{2}\text{H}_{6}
\end{align}
was first proposed by Redfern \textit{et al.} \cite{RZ2000} to show the errors of B3LYP systematically increase as the number of chain units in \textit{n}-alkanes, $m$, increases. 
To assess how PBE0-2 improves upon this problem, we calculate the isodesmic reaction energies of \textit{n}-alkanes to ethane, where the optimized geometries and reference values are taken from Ref.\ \cite{G2010}. 
As shown in Fig.\ \ref{Alkane}, the predicted reaction energies of PBE0-2 are extremely close to the reference values (within an error of 0.02 kcal/mol), due to its accurate treatment of medium-range 
correlation effects \cite{G2010}. By contrast, all the other functionals and MP2 exhibit systematic errors with the increase of $m$. Note that B2PLYP-D3 overcorrects the systematic errors of B2PLYP. 

\begin{figure}
	\includegraphics[scale=0.4]{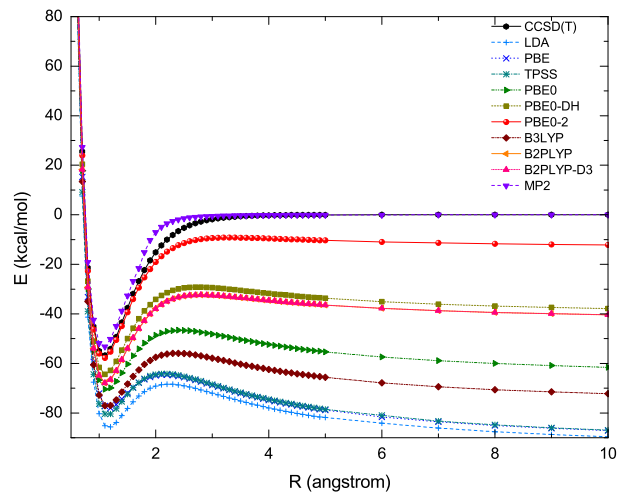}
	\caption{\label{He2+} Dissociation curve of $\text{He}_2^+$. Zero level is set to \textit{E}(He) + \textit{E}($\text{He}^{+}$) for each method.} 
\end{figure}

\begin{figure}
	\includegraphics[scale=0.4]{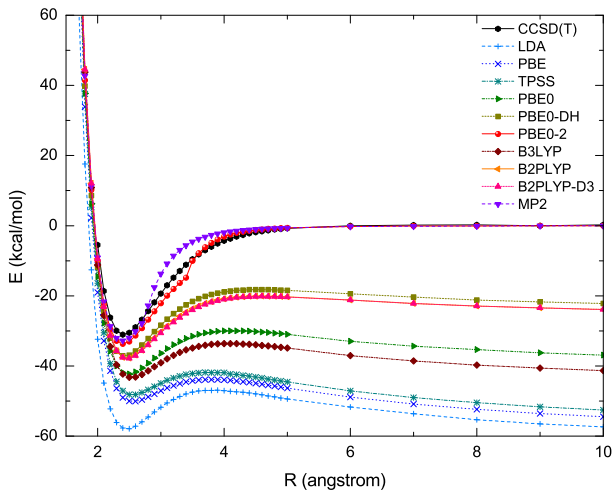}
	\caption{\label{Ar2+} Same as Fig.\ {\ref{He2+}}, but for $\text{Ar}_2^+$.} 
\end{figure}

Due to the severe SIEs of semilocal functionals, spurious fractional charge dissociation can occur, especially for symmetric charged radicals $\text{X}_{2}^{+}$ \cite{SIE}, such as 
$\text{He}_{2}^{+}$ and $\text{Ar}_{2}^{+}$. To examine how PBE0-2 improves upon the SIE problem, spin-unrestricted calculations, using the aug-cc-pVQZ basis set, are performed. The DFT results are compared 
with results from MP2 theory and the highly accurate CCSD(T) theory \cite{RT1989}. As can be seen in Fig.\ \ref{He2+}, the predicted $\text{He}_{2}^{+}$ binding energy curve of PBE0-2 is very close to the CCSD(T) curve. 
It appears that the SIE associated with PBE0-2 is more than three times smaller than the next best DFT method shown. For $\text{Ar}_{2}^{+}$, PBE0-2 can dissociate it correctly (see Fig.\ \ref{Ar2+}), which is a very 
promising result. However, a discontinuity undesirably appears in the derivative of PBE0-2 binding energy curve for $\text{Ar}_2^{+}$, due to the use of MP2 correlation (as discussed in Refs.\ \cite{N-rep,CH2009}). 

\begin{table*}
\footnotesize
\caption{\label{tab1} Mean absolute errors (in kcal/mol) of 10 methods for various test sets (see text for details).}
\begin{ruledtabular}
\begin{tabular}{lcccccccccc}
&  \multicolumn{6}{c}{PBE-based functionals} & \multicolumn{3}{c}{Other functionals} & \multicolumn{1}{c}{\textit{Ab initio}} \\
\cline{2-7}
\cline{8-10}
\cline{11-11}
System & LDA & PBE & TPSS & PBE0 & PBE0-DH & PBE0-2  & B3LYP & B2PLYP & B2PLYP-D3 & MP2\\  \hline 
G3/99 (223) & 121.26 &   21.51 & 5.30 &  6.29 &  5.11 &  5.82  &  5.48 &  3.97 & 2.98 & 9.18\\
IP (40)  &   5.73 &  3.45 &  3.59 &  3.48 &  3.17 &  2.38  &  3.79 &  2.35 & 2.35 & 3.69\\
EA (25)  &   5.03 &  2.41 &  2.61 &  3.10 &  3.55 &  3.01  &  2.38 &  2.16 & 2.17 & 3.95\\
PA (8)  &   5.91 &  1.60 &  1.81 &  1.14 &  1.09 &  0.96  &  1.16 &  1.18 & 1.13 & 0.96\\
NHTBH (38)  &   12.62 & 8.62 &  9.16 &  3.63 &  1.57 &  2.44  &  4.69 &  2.19 & 2.44 & 5.48\\
HTBH (38) &  17.95 & 9.67 &  8.68 &  4.60 &  2.01 &  1.39  &  4.56 &  2.17 & 2.45 & 3.38\\
S22 (22)  & 2.03 & 2.72 & 3.56 & 2.46 & 1.78 	&	0.61 &	3.90 &	2.25 & 0.22 & 0.89\\
S66 (66)  &   1.85 &  2.23 &  3.09 &  2.10 &  1.58 &  0.67  &  3.33 &  1.96 & 0.21 & 0.45\\
\end{tabular}
\end{ruledtabular}
\end{table*}

\section{Conclusions}

In summary, we have developed a PBE-based double-hybrid functional, employing a half-and-half mixing of the PBE and MP2 correlation functionals with the LS1DH approximation in Eq.\ (\ref{LS1DH}). 
Owing to its significant improvement over LDA, PBE, TPSS, and PBE0, this functional, denominated PBE0-2, fits well into the fifth rung of the ladder. Our results indicate that PBE0-2 is generally comparable or 
superior to PBE0-DH in performance. As PBE0-2 contains a very large fraction ($\approx$ 79 \%) of HF exchange and a large fraction (50 \%) of MP2 correlation, the SIE and NCIE associated with PBE0-2 
are significantly reduced, reflecting its superior performance for self-interaction problems and noncovalent interactions, respectively. However, due to the perturbative treatment of nonlocal correlation, 
the SCE of PBE0-2 can be enormous in situations where strong static correlation effects are pronounced \cite{DFTreview,Chai2012}. It remains challenging to develop a generally accurate density functional 
resolving all the qualitative failures of semilocal approximations at a reasonable computational cost.

\begin{acknowledgments}

This work was supported by National Science Council of Taiwan (Grant No. NSC98-2112-M-002-023-MY3), National Taiwan University (Grant Nos. 99R70304 and 10R80914-1), and NCTS of Taiwan. 

\end{acknowledgments}

\bibliographystyle{cpl}

\begin{references}

\bibitem{KS1965} W. Kohn, L.J. Sham, Phys. Rev. 140 (1965) A1133. 
\bibitem{DFTreview} A.J. Cohen, P. Mori-S\'anchez, W. Yang, Chem. Rev. 112 (2011) 289. 
\bibitem{constraint} J.P. Perdew, A. Ruzsinszky, J. Tao, V.N. Staroverov, G.E. Scuseria, G.I. Csonka, J. Chem. Phys. 123 (2005) 062201. 
\bibitem{LDAX} P.A.M. Dirac, Proc. Cambridge Philos. Soc. 26 (1930) 376. 
\bibitem{LDAC} J.P. Perdew, Y. Wang, Phys. Rev. B 45 (1992) 13244.
\bibitem{PB1996} J.P. Perdew, K. Burke, M. Ernzerhof, Phys. Rev. Lett. 77 (1996) 3865.
\bibitem{TP2003} J. Tao, J.P. Perdew, V.N. Staroverov, G.E. Scuseria, Phys. Rev. Lett. 91 (2003) 146401.
\bibitem{Chai2012} J.-D. Chai, J. Chem. Phys. 136 (2012) 154104. 
\bibitem{B1993_2} A.D. Becke, J. Chem. Phys. 98 (1993) 5648.
\bibitem{DFA0} J.P. Perdew, M. Ernzerhof, K. Burke, J. Chem. Phys. 105 (1996) 9982. 
\bibitem{PBE0} C. Adamo, V. Barone, J. Chem. Phys. 110 (1999) 6158. 
\bibitem{DH1} Y. Zhao, B.J. Lynch, D.G. Truhlar, J. Phys. Chem. A 108 (2004) 4786. 
\bibitem{DH2} Y. Zhao, B.J. Lynch, D.G. Truhlar, Phys. Chem. Chem. Phys. 7 (2005) 43. 
\bibitem{G2006} S. Grimme, J. Chem. Phys. 124 (2006) 034108. 
\bibitem{XYG3} Y. Zhang, X. Xu, W.A. Goddard III, Proc. Natl. Acad. Sci. U.S.A. 106 (2009) 4963 .
\bibitem{CH2009} J.-D. Chai, M. Head-Gordon, J. Chem. Phys. 131 (2009) 174105. 
\bibitem{B2PLYPD3} S. Grimme, J. Antony, S. Ehrlich, H. Krieg, J. Chem. Phys. 132 (2010) 154104. 
\bibitem{MP1934} C. M\o ller, M.S. Plesset, Phys. Rev. 46 (1934) 618.
\bibitem{ST2011} K. Sharkas, J. Toulouse, A. Savin, J. Chem. Phys. 134 (2011) 064113.
\bibitem{BA2011} E. Br\'emond, C. Adamo, J. Chem. Phys. 135 (2011) 024106. 
\bibitem{TS2011} J. Toulouse, K. Sharkas, E. Br\'emond, C. Adamo, J. Chem. Phys. 135 (2011) 101102. 
\bibitem{Dobson2} J.F. Dobson, K. McLennan, A. Rubio, J. Wang, T. Gould, H.M. Le, B.P. Dinte, Aust. J. Chem. 54 (2001) 513. 
\bibitem{DFT-D1} S. Grimme, J. Comput. Chem. 25 (2004) 1463. 
\bibitem{DFT-D2} S. Grimme, J. Comput. Chem. 27 (2006) 1787. 
\bibitem{wB97X-D} J.-D. Chai, M. Head-Gordon, Phys. Chem. Chem. Phys. 10 (2008) 6615. 
\bibitem{vdW} M. Dion, H. Rydberg, E. Schr\"oder, D.C. Langreth, B.I. Lundqvist, Phys. Rev. Lett. 92 (2004) 246401.
\bibitem{Sherrill} L.A. Burns, \'A.V\'azquez-Mayagoitia, B.G. Sumpter, C.D. Sherrill, J. Chem. Phys. 134 (2011) 084107. 
\bibitem{B3LYP} P.J. Stephens, F.J. Devlin, C.F. Chabalowski, M.J. Frisch, J. Phys. Chem. 98 (1994) 11623.
\bibitem{CR1997} L.A. Curtiss, K. Raghavachari, P.C. Redfern, J.A. Pople, J. Chem. Phys. 106 (1997) 1063. 
\bibitem{CR1998} L.A. Curtiss, P.C. Redfern, K. Raghavachari, J.A. Pople, J. Chem. Phys. 109 (1998) 42. 
\bibitem{CR2000} L.A. Curtiss, K. Raghavachari, P.C. Redfern, J.A. Pople, J. Chem. Phys. 112 (2000) 7374.
\bibitem{PH1989} J.A. Pople, M. Head-Gordon, D.J. Fox, K. Raghavachari, L.A. Curtiss, J. Chem. Phys. 90 (1989) 5622. 
\bibitem{ZL2004_2} Y. Zhao, B.J. Lynch, D.G. Truhlar, J. Phys. Chem. A 108 (2004) 2715. 
\bibitem{ZG2005} Y. Zhao, N. Gonz\'alez-Garc\'ia, D.G. Truhlar, J. Phys. Chem. A 109 (2005) 2012; 110 (2006) 4942(E).
\bibitem{JS2006} P. Jure\v{c}ka, J. \v{S}poner, J. \v{C}ern\'y, P. Hobza, Phys. Chem. Chem. Phys. 8 (2006) 1985. 
\bibitem{TH2010} T. Takatani, E.G. Hohenstein, M. Malagoli, M.S. Marshall, C.D. Sherrill, J. Chem. Phys. 132 (2010) 144104.
\bibitem{RL2011} J. Rezac, K.E. Riley, P. Hobza, J. Chem. Theory Comput. 7 (2011) 2427.
\bibitem{SF2006} Y. Shao {\it et al.}, Phys. Chem. Chem. Phys. 8 (2006) 3172.
\bibitem{wB97} J.-D. Chai, M. Head-Gordon, J. Chem. Phys. 128 (2008) 084106. 
\bibitem{KF1997} R.A. Kendall, H.A. Fr\"uchtl, Theor. Chem. Acc. 97 (1997) 158. 
\bibitem{BB1970} S.F. Boys, F. Bernardi, Mol. Phys. 19 (1970) 553.
\bibitem{supp} See Supplementary Material Document No. (to be inserted) for further numerical results. 
\bibitem{RZ2000} P.C. Redfern, P. Zapol, L.A. Curtiss, K. Raghavachari, J. Phys. Chem. A 104 (2000) 5850.
\bibitem{SE1} M.D. Wodrich, C. Corminboeuf, P.v.R. Schleyer, Org. Lett. 8 (2006) 3631. 
\bibitem{ST2010} J.-W. Song, T. Tsuneda, T. Sato, K. Hirao, Org. Lett. 12 (2010) 1440. 
\bibitem{G2010} S. Grimme, Org. Lett. 12 (2010) 4670. 
\bibitem{SE2} S.N. Steinmann, M.D. Wodrich, C. Corminboeuf, Theor. Chem. Acc. 127 (2010) 429. 
\bibitem{SIE} T. Bally, G.N. Sastry, J. Phys. Chem. A 101 (1997) 7923. 
\bibitem{RT1989} K. Raghavachari, G.W. Trucks, J.A. Pople, M. Head-Gordon, Chem. Phys. Lett. 157 (1989) 479. 
\bibitem{N-rep} W. Kurlancheek, M. Head-Gordon, Mol. Phys. 107 (2009) 1223. 

\end{references}

\end{document}